\documentclass[twocolumn]{svjour3}

\smartqed

\usepackage{graphicx}
\usepackage{bm}
\usepackage{xcolor}

\begin{document}

\setcounter{page}{1}

\title{Thermo-elasto-plastic simulations of femtosecond laser-induced multiple-cavity in fused silica}

\author{R. Beuton \and B. Chimier \and J. Breil \and D. H\'ebert \and K. Mishchik \and J. Lopez \and P. H. Maire \and G. Duchateau}

\institute{R. Beuton \and B. Chimier \and J. Lopez \and G. Duchateau \at 
	   Universit\'e de Bordeaux-CNRS-CEA, Centre Lasers Intenses et Applications, UMR 5107, 33405 Talence, France\\
	   \email{romain.beuton@u-bordeaux.fr}
           \and
           J. Breil \and D. H\'ebert \and P. H. Maire \at
           CEA/CESTA, 15 Avenue des Sabli\`eres, CS 60001 33116 Le Barp cedex France
           \and
           K. Mishchik \at
           Amplitude Systemes, Pessac, France \\
           Universit\'e de Bordeaux-CNRS-CEA, Centre Lasers Intenses et Applications, UMR 5107, 33405 Talence, France
}

\date{Received: date / Accepted: date}
 
\maketitle 
 
\begin{abstract}
The formation and the interaction of multiple cavities, induced by tightly focused femtosecond laser pulses, are studied by using a developed numerical tool, including the thermo-elasto-plastic material response. Simulations are performed in fused silica 
in cases of one, two, and four spots of laser energy deposition. The relaxation of the heated matter, launching shock waves in the surrounding cold material, leads to cavity formation and emergence of areas where cracks may be induced.
Results show that the laser-induced structure shape depends on the energy deposition configuration and demonstrate the potential of the used numerical tool to obtain the desired 
designed structure or technological process.
\end{abstract}

\section{INTRODUCTION}
Modifying the properties of dielectric materials, by inducing designed structures on the surface or in the bulk,
is interesting for various industrial and technological applications \cite{Davis1996,Shimotsuma2003,Juodkazis2003,Gattass2008,Gottmann2009,Cheng2009,Bonse2012,Richter2012,Kumkar2014,Buividas2014,Liao2015,Hendricks2016}. 
Due to the nonlinear mechanisms of the laser energy
absorption and the low thermal diffusion during the laser irradiation, femtosecond laser pulses are an attractive tool for microprocessing such kind of materials.

By using tightly focused femtosecond laser pulses, 3D nano-structures such as cavities can be induced in the bulk of dielectrics \cite{Glezer1997,Juodkazis2006,Gamaly2006,Hallo2007,Mezel2008,Mishchik2010,Vailionis2011,Bhuyan2017}.
During the laser-matter interaction, nonlinear processes lead to a significant absorption of the laser energy in the focal volume.
The heated material is confined, creating a warm dense matter surrounded by a non-modified solid cold material.
The induced high pressure and temperature, in the heated volume, then launche a shock wave which propagates in the surrounding matter, followed by rarefaction waves which dig the material.
Plastic and elastic deformations are then induced in the surrounding matter depending on the irradiation conditions.
For absorbed energy densities above the material damage threshold, a significant permanent structural modification may appear due to the generated plastic deformations induced by the compression and traction waves 
in the solid matter \cite{Sakakura2007,Hebert2011,Bulgakova2015,Najafi2016}.

To understand the  physical processes taking place during the heated volume relaxation, in
the particular case of tightly focused beams, a thermo-elasto-plastic
model has been recently implemented in a lagrangian hydrodynamic code \cite{Beuton2017}. 
A model for the predictions of the critical areas, where cracks may appears, has been also presented.
Results have shown the importance of
the solid response (the elasto-plastic behavior) in the cavity generation process and the effect of the energy deposition geometry on the potential cracks formation.

In the present paper, this previous work and the developed numerical tools are used to simulate the generation of multiple-cavity structures in the bulk of fused silica. 
A short description of the theoretical model \cite{Wilkins1964,Maire2013} is presented in Section \ref{th.modeling}.
The simulations are performed by assuming an instantaneous energy deposition.
The single-cavity case is first considered, as it corresponds 
to the simplest configuration, to understand 
the main phenomena at play. It is then generalized to the study of double- and multiple-cavity generation, where the interaction between sub-structure is investigated. Especially, various cases will be studied to highlight 
the cooperative mechanisms responsible for the formation of particular nano-structures and micro-structures, useful in many applications \cite{Courvoisier2013,Sakakura2013,Courvoisier2016,Bellouard2016}.
The case of two cavities, separated by a few tenth of micrometers, shows an increase of the compressive and shear stresses, compared to the single cavity generation, which leads to critical areas by compression during a time  larger than the single
spot one. This case also induces critical areas associated to shear.
Furthermore, the influence of the distance between these two cavities is studied. Finally, the interaction between multiple cavities is presented in two different cases (simultaneous and delayed energy depositions). 
Different shapes of the formed cavities and spatial repartitions of the critical areas are obtained, depending on the considered case.
The results and discussions, for the different studied cases, are presented in Section \ref{results}.
Conclusions and perspectives are drawn in Section \ref{conclusion}.

\section{THERMO-ELASTO-PLASTIC MODEL}
\label{th.modeling}
The hydrodynamic response of the material, locally submitted to a fast heating, can be described by a generalized lagrangian form of the hydrodynamic conservation laws \cite{Maire2013,Beuton2017}, 
where the thermodynamic pressure is substituted by the Cauchy
stress tensor $\bar{\bar\sigma}$ \cite{Irgens2008}: 
\begin{equation}
 \rho\frac{d}{dt}(\frac{1}{\rho})-\nabla.\textbf{V}=0
\end{equation}

\begin{equation}
 \rho\frac{d\textbf{V}}{dt}-\nabla.\bar{\bar\sigma}=\textbf{0}
\end{equation}

\begin{equation}
 \rho\frac{dE}{dt}-\nabla.(\bar{\bar\sigma}.\textbf{V})=0
\end{equation}
$\rho$, $\textbf{V}$ and $E$ are the density, the velocity and the specific total energy, respectively.

The Cauchy stress tensor can be decomposed into two terms:
\begin{equation}
 \bar{\bar\sigma}=-PI_d+\bar{\bar S} 
 \label{sigma}
\end{equation}
where the first term, depending on the thermodynamic pressure $P$, accounts for the fluid response of the matter ($I_d$ is the identity matrix). The second term, depending on the deviatoric part of the Cauchy stress
tensor $\bar{\bar S}$, accounts for the solid behavior, especially by including the shear stresses. 

The thermodynamic pressure is evaluated through an equation of state (EOS), while $\bar{\bar S}$ is evaluated by solving a
differential equation depending on the shear modulus, the strain rate tensor and the antisymmetric part of the velocity gradient \cite{Maire2013}. Since the strain rate tensor depends on the regime of deformation, a yield criterion is required. 
In the present study, the von Mises yield criterion is used \cite{Mises1913}.
In this case, an equivalent stress can be defined as:
 \begin{equation}
 \sigma_{eq} =\sqrt{\frac{3}{2}Tr(\bar{\bar S}.\bar{\bar S})}
\end{equation}
By comparing this local effective stress to the yield strength of the material, \textit{i.e.} its elastic limit, the regime of deformation is deduced. The plastic regime takes place for equivalent stresses exceeding the yield strength, while
the elastic regime occurs for equivalent stresses lower than the yield strength \cite{Beuton2017}.

From the knowledge of the Cauchy stress tensor, the present thermo-elasto-plastic model allows one to perform
the laser induced principal stresses in compression-traction and the maximum shear stresses, in each spatial position.
That permits to determine different critical areas where potential cracks (fractures) may appear by comparing them to the intrinsic mechanical limits.
The strength limits of the material can be defined by the resistance 
in compression $L_c$, traction $L_t$ and shear $L_t/2$ (the shear stress is the most efficient mechanism to induce cracks in materials) \cite{Spenle2003}.
If the principal stresses or maximum shear stress exceeds the associated previous limits, then the material may break, \textit{i.e.} 
cracks may appear \cite{Beuton2017}.

Finally, a thermal softening is introduced in the model to account for the solid-liquid phase transitions, depending on the specific internal energy. 
For an internal energy larger than the energy of melting, the solid behavior is removed ($\bar{\bar S}$ is canceled) and the classical Euler's equations are retrieved.

Based on this model, a second order cell-centered Lagrangian numerical scheme has been developed in a 2D planar geometry \cite{Maire2013}. It has been implemented in a hydrodynamic code \cite{Breil2010} to study the formation of cavity,
induced by tightly focused femtosecond laser pulse, in dielectrics \cite{Beuton2017}.

\section{RESULTS AND DISCUSSION}
\label{results}
In the present study, the formation of different nanostructures, based on the generation of cavities in the bulk of fused silica, are considered. For each energy deposition, 
a gaussian shape is assumed with a radius set to 0.13 $\mu$m at half maximum \cite{Gamaly2006} (such a small size for the energy deposition is possible due to nonlinear effects).
A single absorbed energy is set to 6 nJ to get a final cavity radius around 0.4 $\mu$m, according to the experimental observations. 
Note that a change in the amount of energy deposition leaves unchanged the general trends provided hereafter, the main influence being to modify as usually the amplitude of waves, their dynamics, etc...
The EOS used in the simulations, giving the pressure and the temperature from the knowledge of the density and the specific internal energy, is the SESAME table 7386 \cite{Boettger1989} for fused silica.
Within these conditions, the EOS predicts initial pressure and temperature around 0.3 TPa and $10^5$ K, respectively, for each energy deposition area.
The initial density is equal to 2.2 g/cm$^3$.
The yield strength and the shear modulus, for initial standard conditions, are $Y_0=7.1$ GPa and $\mu_0=22.6$ GPa, respectively \cite{Beuton2017}.
The strength limit in traction $L_t$ and in compression $L_c$ are set to characteristic values (microscopic scale) of 8 GPa and 10 GPa, respectively \cite{Pedone2015} (resistance in compression 
is generally larger than resistance in traction for brittle materials as fused silica\cite{Fanchon2001}). 
The full cartesian mesh size for simulations is set to 18 $\mu$m$\times$18 $\mu$m, with initial cell size set to 50 nm$\times$50 nm. 
All the previous parameters are used throughout the paper.

\subsection{Single energy deposition}
\label{single}
Here is considered a single energy deposition in the center of the 2D domain.

\begin{figure}[!h]
  \centering
  \includegraphics[width=8cm]{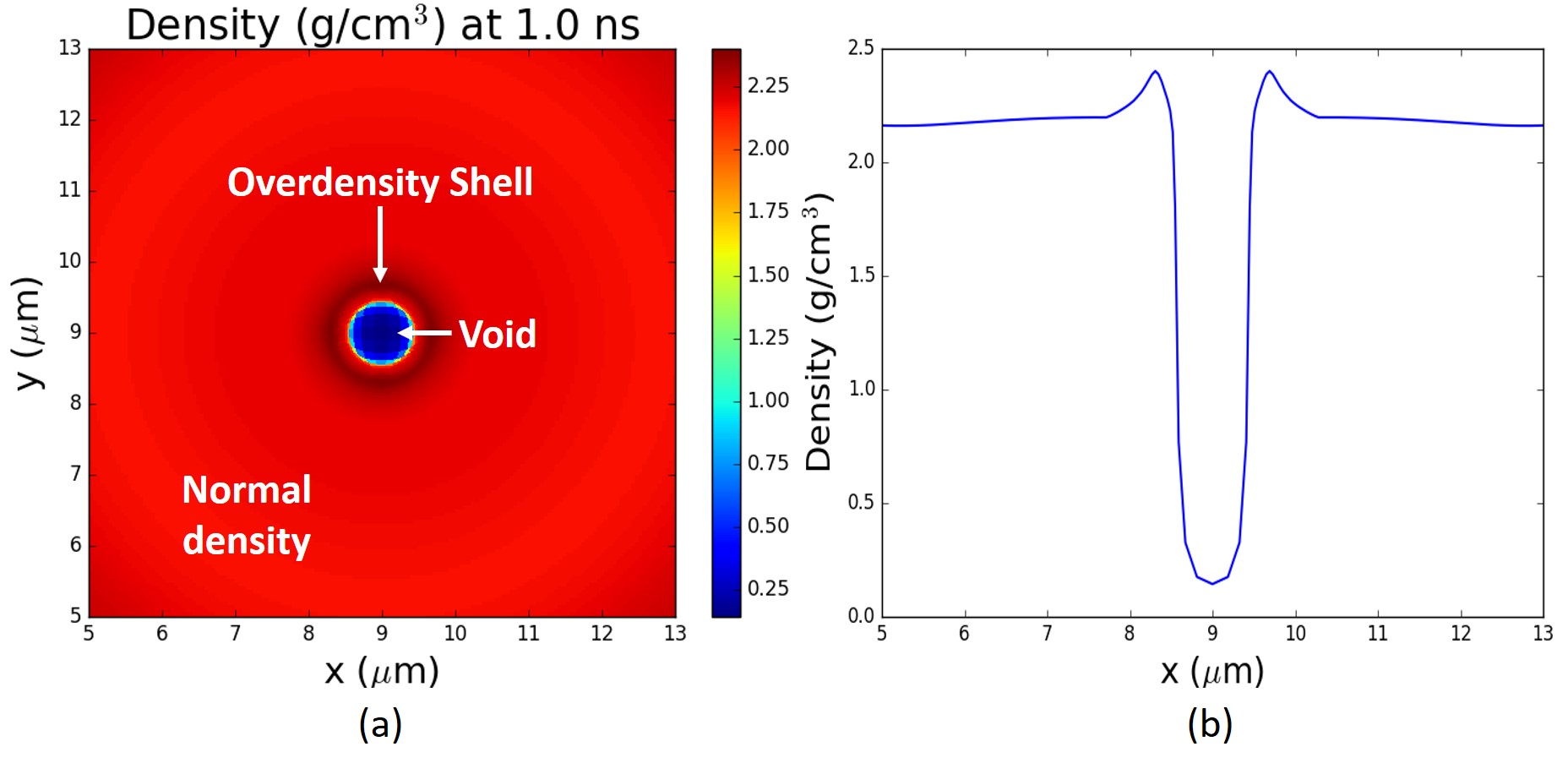}
  \caption{(a) 2D density map and (b) associated radial density profile at 1 ns.}
  \label{Figure_1}
\end{figure}
Figure \ref{Figure_1} presents the 2D spatial map of the density (Fig. \ref{Figure_1}(a)) and its associated radial density profile (Fig. \ref{Figure_1}(b)), given by the simulation at 1 ns after the energy deposition. 
Within these conditions, a cavity with a radius 
around 0.4 $\mu$m is formed, surrounded
by an overdensity shell, stabilized at this time. The void and the overdensity have a density lower than 0.25 g/cm$^3$ and larger than 2.25 g/cm$^3$, respectively.

\begin{figure}[!h]
  \centering
  \includegraphics[width=8cm]{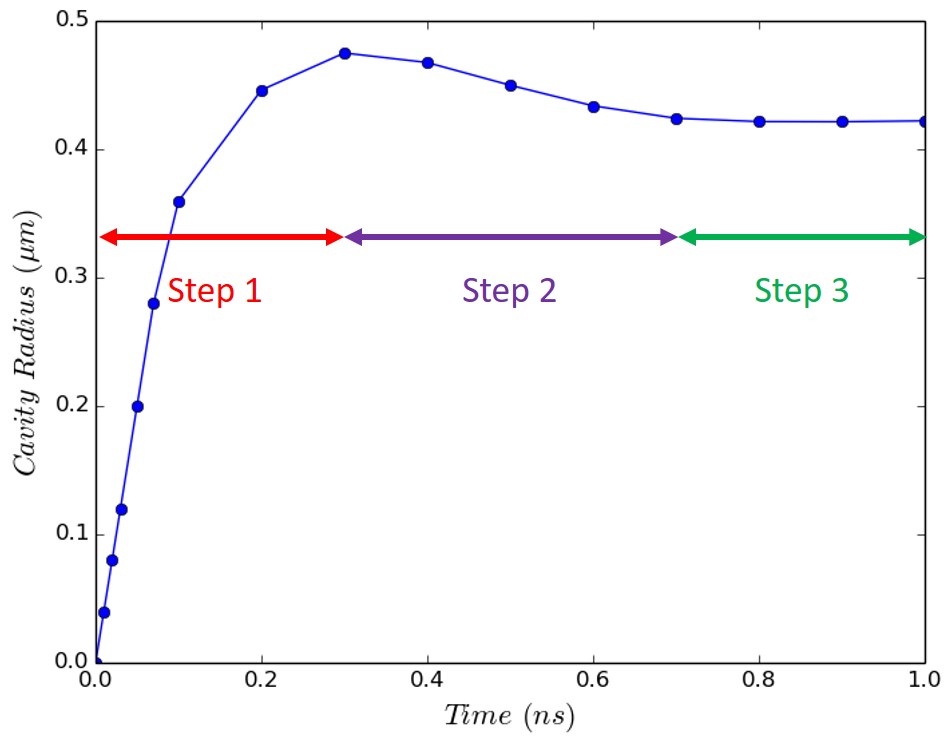}
  \caption{Evolution of the cavity radius as a function of time. It can be decomposed into three steps as shown by the horizontal arrows.}
  \label{Figure_2}
\end{figure}
The temporal evolution of the cavity radius is presented in Fig. \ref{Figure_2}. It is mainly composed of three steps: an expansion, a contraction (shrinking) and a stabilization \cite{Beuton2017}.
The first step is due to the formation of a strong shock, followed by rarefaction waves which form the void. In the same time, plastic and elastic deformations are induced (elasto-plastic wave) in the surrounding material 
and stresses are accumulated. When the shock transforms into a pure elastic wave (stresses has decreased after the shock), the expansion stops and the cavity radius reaches its maximum value 
(which does not correspond to the maximum stressed state). 
Then, corresponding to the above-mentioned second step, 
the decompression of the previously compressed material by the shock, \textit{i.e.} the relaxation of the elastic deformations, close partially the cavity. 
Finally, the cavity stabilizes (third step) due to the previously induced permanent deformations (plastic deformations).

Within these conditions, an evaluation of the principal stresses has been performed. Maximum stresses are observed 200 ps after the energy deposition, during the expansion of the cavity.
\begin{figure}[!h]
  \centering
  \includegraphics[width=8cm]{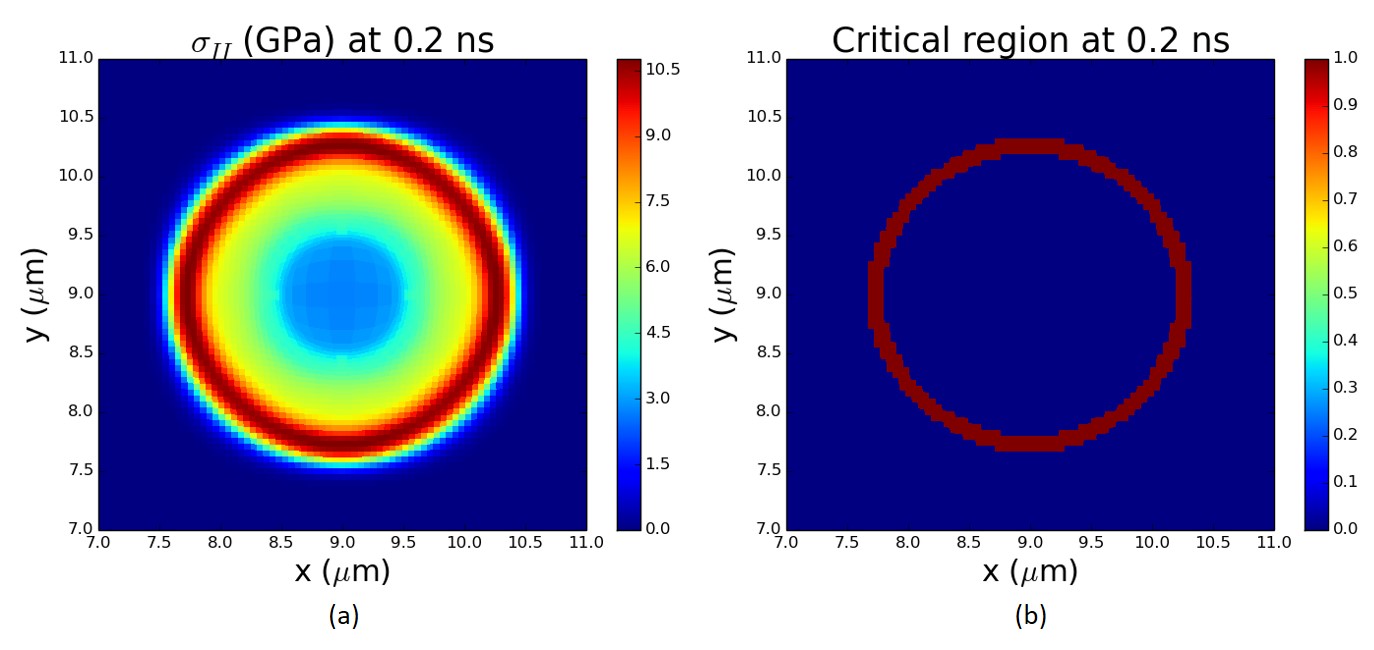}
  \caption{(a) 2D map of principal stresses in compression $\sigma_{II}$. (b) Critical area in compression around the cavity.}
  \label{Figure_3}
\end{figure}
However, only compressive stresses can induce a critical area,
where their values are greater than the strength limit in compression $L_c$. Shear stresses are not high enough to induce a critical zone due to the homogeneous distribution of the compressive wave in the present case.
Principal stress in compression $\sigma_{II}$, with its associated critical area, are presented in Fig. \ref{Figure_3}.
Due to the isotropic energy deposition, the compression is homogeneously distributed
around the created void (Fig. \ref{Figure_3}(a)) and the associated critical region is a ring (Fig. \ref{Figure_3}(b)). Finally, this critical region disappears around 220 ps, when the amplitude of the shock wave is lower than $L_c$.

\subsection{Double energy deposition}
\label{double}
The case of a simultaneous double energy deposition is now considered. The total energy is set to 12 nJ, which is twice the previous absorbed energy,
distributed within two main spots separated by 0.72 $\mu$m. In this case, two cavities, separated by a slightly denser region, are created. The interaction between the two shock waves, launched by each energy deposition, 
leads to a particular behavior. In the horizontal direction outwards the cavities, the shock waves mainly propagate as in the previous single cavity case. In the inward direction, the shock waves collide, leading to a 
daughter wave in the perpendicular direction with a larger amplitude.

\begin{figure}[!h]
  \centering
  \includegraphics[width=8cm]{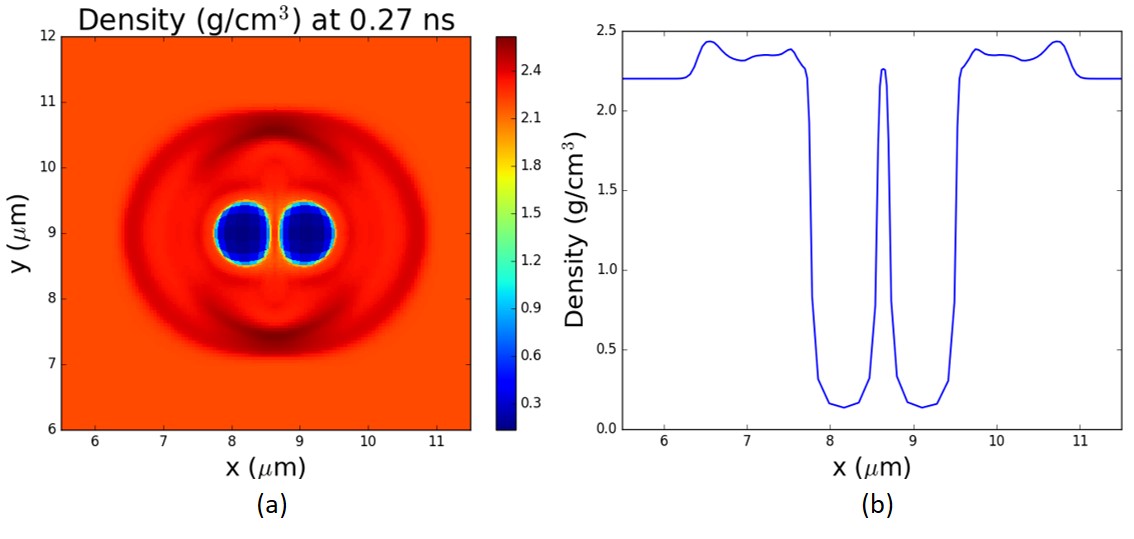}
  \caption{(a) 2D map of the density and (b) the density profile along the x axis at 250 ps.}
  \label{Figure_4}
\end{figure}
Figures \ref{Figure_4} presents the 2D density map and the density profile along the x axis, at 270 ps after the energy deposition. The two formed cavities are surrounded by an inhomogeneous overdense zone, where
two symmetric regions are highly compressed in the vertical direction due to the daughter wave. Moreover, the asymmetric propagation and shape of the shock waves lead to an elongation of the two cavities
in the vertical direction, as observed experimentally by Zimmermann  \textit{et al.} \cite{Zimmermann2016}. In this case, shear stresses may be high enough to induce potential cracks. 

\begin{figure}[!h]
  \centering
  \includegraphics[width=8cm]{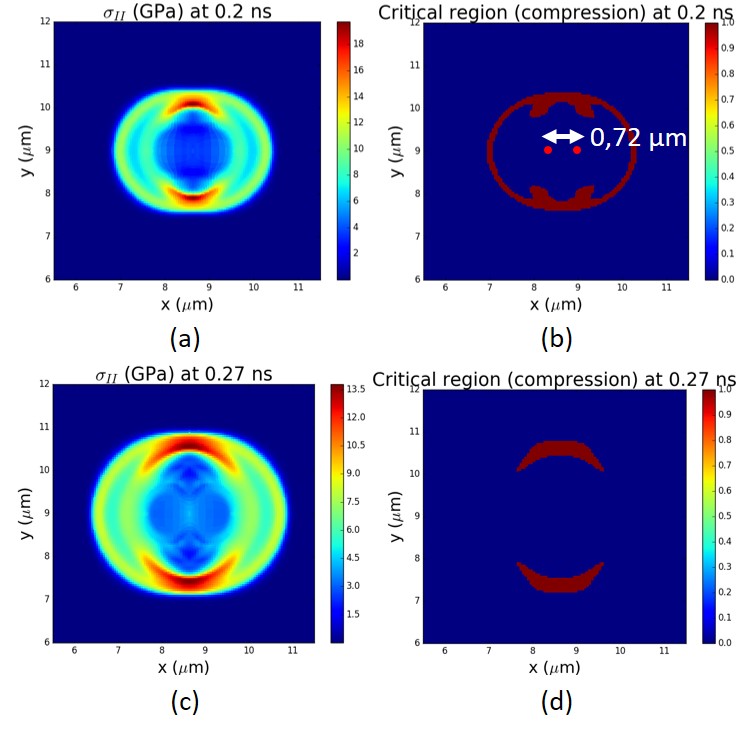}
  \caption{2D spatial profiles of the principal stresses in compression with their associated critical areas at (a,b) 200 ps and (c,d) 270 ps.}
  \label{Figure_5}
\end{figure}

\begin{figure}[!h]
  \centering
  \includegraphics[width=8cm]{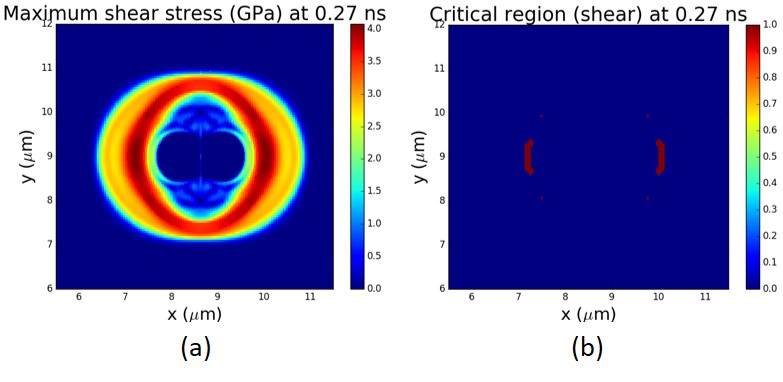}
  \caption{2D spatial profiles of (a) the maximum shear stresses with (b) its associated critical area at 250 ps.}
  \label{Figure_6}
\end{figure}

The evaluation of the principal stresses in compression 
and the maximal shear stresses, with their associated critical areas, are presented in Figs. \ref{Figure_5} and \ref{Figure_6}, respectively. Two different times are considered for the principal stresses in compression.
At 200 ps after the energy deposition (Figs. \ref{Figure_5}(a) and (b)), the contributions of both the mother and the daughter waves are visible, leading to a critical area by compression in Fig. \ref{Figure_5}(b) relatively isotropic, as 
in Fig. \ref{Figure_3}(b). However, as discussed in the previous section, the mother wave amplitude is lower than $L_c$ after around 220 ps and no longer contributes to
induce potential cracks.
At 270 ps after the energy deposition (Figs. \ref{Figure_5}(c) and (d)), only the contribution of the daughter wave induces a high enough compressive stress in the vertical direction, resulting in both the upper and lower regions 
of maximal stress in excess of the strength limit, as shown in Fig. \ref{Figure_5}(d). 
Moreover, the asymmetric distribution of compressive stresses, at this time, leads to an elongation of the cavities in the vertical direction. 
It induces strong enough shear stresses (maximal at 270 ps) on the left and right sides (Fig. \ref{Figure_6}(a)) to create two symmetric associated critical areas in the horizontal direction, as depicted by Fig. \ref{Figure_6}(b).

The evolution of the critical regions depends on the distance $d$ between the two centers of the energy deposition spots. If the two spots are too close ($d\leq0.43$ $\mu$m within the present conditions), 
the two generated cavities merge during their growth, ultimately leading to only one cavity. The distribution
of the compressive stresses, in the surrounding matter, is almost homogeneous as in Fig. \ref{Figure_3}. The amplitude of the shear stresses becomes too low to induce any associated critical area. If the distance $d$ is too large
($d\geq0.81$ $\mu$m), the amplitude of the daughter wave, resulting from the two shock waves collision, is not high enough to lead to an important inhomogeneous repartition of compressive stresses. The resulting
shear stress is then not high enough to induce any critical area and potential cracks.

\subsection{Application to laser-induced multiple cavities}
The double cavities are now generalized to the multiple-cavity case. The formation of four cavities is investigated here, this number being sufficiently large to exhibit complex waves interaction.
The total energy is set to 24 nJ and the distance between two neighbor spots is 0.72 $\mu$m.
In a first case, four simultaneous energy depositions are considered. Four shock waves are then launched simultaneously. Their interaction leads to various daughter waves, with a larger amplitude than the double-cavity case, 
propagating in the vertical direction.
In a second case, the four energy depositions are absorbed with a delay of 100 ps between two subsequent spots. This delay value ensures that each energy deposition takes place in the non-modified material. 
In this case, the four shock waves, which are launched at different times, interact mainly in the horizontal direction with an accumulation of compression waves in this direction. In the vertical direction, the formation of daughter waves cannot 
appear due to the delayed launching of the shock waves.

\begin{figure}[!h]
  \centering
  \includegraphics[width=8cm]{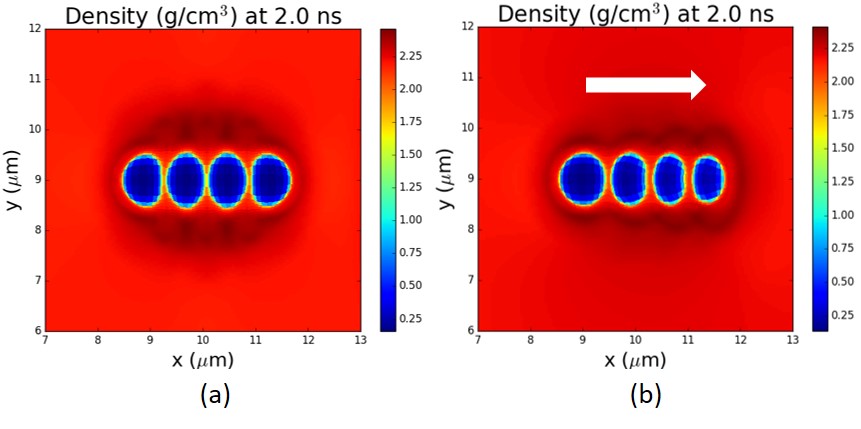}
  \caption{2D maps of the final density profiles in the case of (a) the simultaneous and (b) the delayed energy depositions.}
  \label{Figure_7}
\end{figure}

Figure \ref{Figure_7} shows the 2D spatial maps of the final density profiles in the case of the simultaneous (Fig. \ref{Figure_7}(a)) and the delayed (Fig. \ref{Figure_7}(b)) four energy depositions.
In the first case (Fig. \ref{Figure_7}(a)), four elongated cavities are formed, with roughly the same size and the same aspect ratio. They are surrounded by an inhomogeneous overdensity area, larger in the vertical direction. As explained previously,
the interacting shock waves induce daughter waves with a larger amplitude
in the vertical axis. Thus the matter is more strongly compressed in this direction, leading to the observed overdensity zone. 

In the second case (Fig. \ref{Figure_7}(b)), four elongated cavities are also generated (from the left to the right as indicated 
by the arrow). 
However, their sizes and aspect ratios are slightly
different. The first cavity (on the left), which expand freely during 100 ps, is roughly similar to the single energy deposition case, but the last cavity (on the right) is smaller and more elongated in the vertical direction. 
This is due to the progressive accumulation of shock waves, leading to a stressed state more important for each new forming cavity. Indeed, 
the expansion is more strongly slowed down because each new cavity is submitted to the shock waves of the previous energy depositions. Furthermore, for the same reasons (accumulation of compressive stresses), 
the overdensity regions, around each cavity, slightly increase from the left
to the right.

\begin{figure}[!h]
  \centering
  \includegraphics[width=8cm]{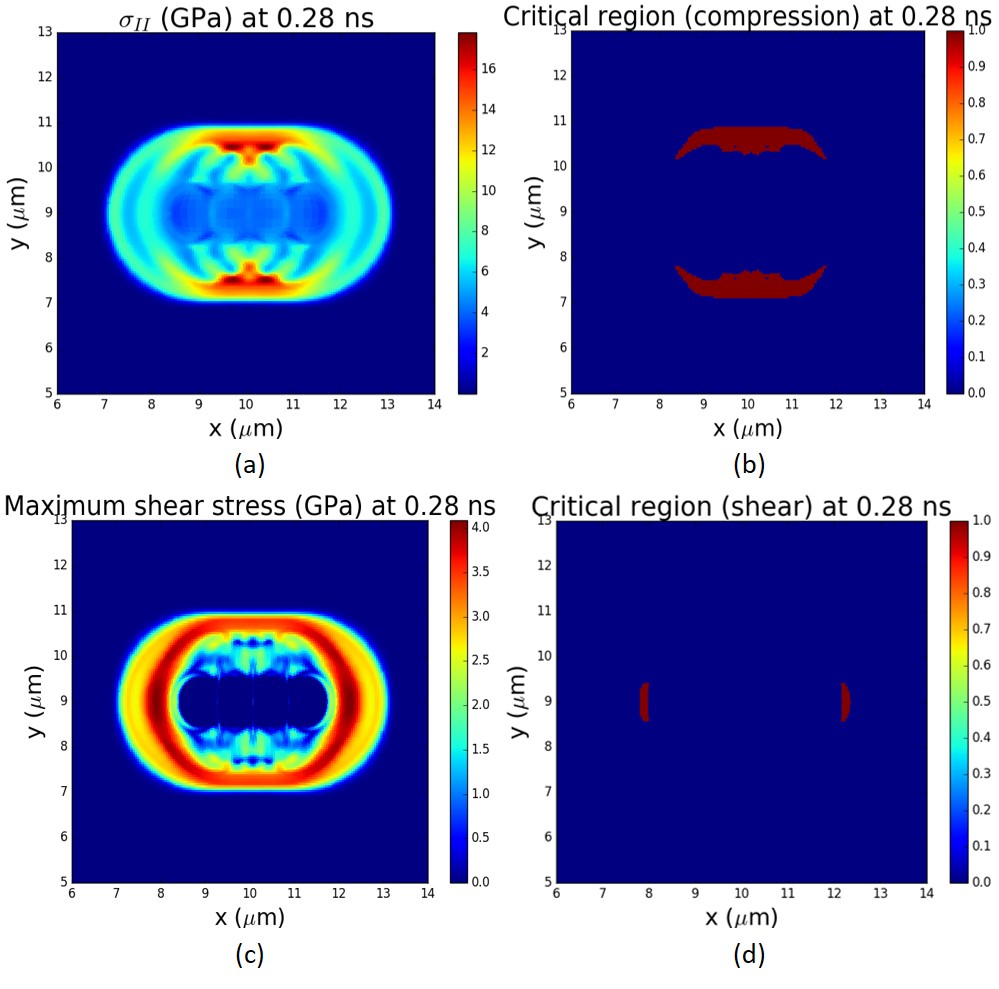}
  \caption{In the case of the simultaneous energy depositions: 2D spatial profiles at 280 ps for (a) the principal stress in compression with (b) its associated critical area; (c) the maximum shear stress with (d) its associated critical area.}
  \label{Figure_8}
\end{figure}
Figure \ref{Figure_8} shows the 2D maps of the evaluated principal stresses in compression (Fig. \ref{Figure_8}(a)) and the maximum shear stresses (Fig. \ref{Figure_8}(c)), with their associated critical areas (Fig. \ref{Figure_8}(b) and (d)), 
in the case of the simultaneous energy depositions, at 280 ps after the energy deposition (time exhibiting the largest shear stress).
As in the double cavities case, the formation of daughter waves leads to two symmetric highly compressed zone in the vertical direction (Fig. \ref{Figure_8}(a)), inducing a large critical regions by compression (Fig. \ref{Figure_8}(b)). 
This is followed by an increase of the shear stresses in the horizontal direction (Fig. \ref{Figure_8}(c)), leading to two critical areas induced by shear (Fig. \ref{Figure_8}(d)). No shear stresses appear between the cavities due 
to the presence of softened glass
removing the solid behavior. An increase of the distance between the energy deposition spots should conserve more solid matter between the cavities and should lead to the apparition of shear stresses in these regions. 
However, as seen in the previous section, for too large distances ($d\geq0.81$ $\mu$m within the present conditions), the effect of the daughter waves is not important enough to induce shear stresses greater than the material resistance in shear
$L_t/2$ and potential cracks between the cavities cannot appear.

\begin{figure}[!h]
  \centering
  \includegraphics[width=8cm]{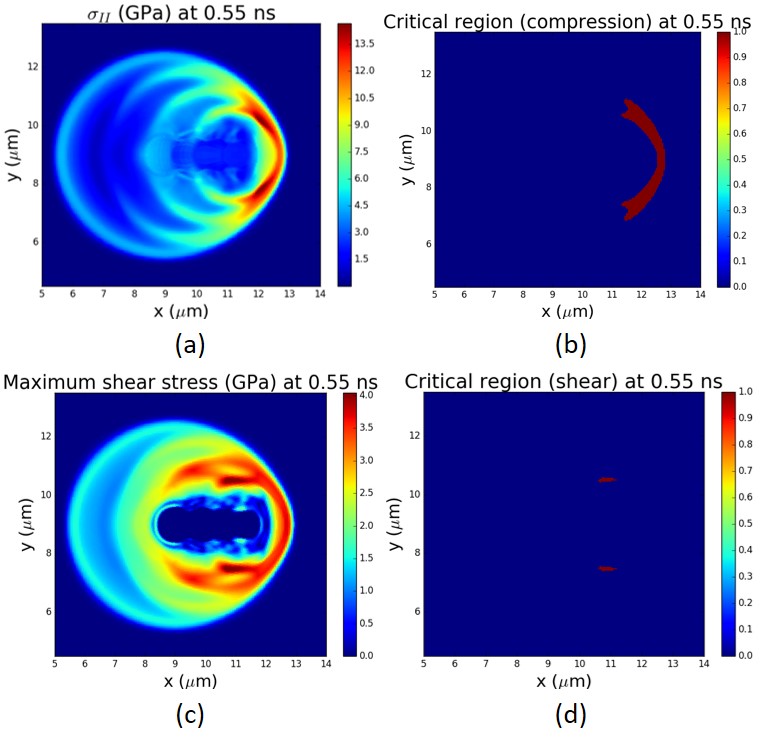}
  \caption{In the case of the delayed energy depositions: 2D spatial profiles at 550 ps for (a) the principal stress in compression with (b) its associated critical area; (c) the maximum shear stress with (d) its associated critical area.}
  \label{Figure_9}
\end{figure}
Figure \ref{Figure_9} presents the principal stresses in compression and the maximum shear stresses, with their associated critical area, for the case of
the delayed energy depositions, at 550 ps after the first energy deposition (time exhibiting the largest shear stress).
The accumulated compression waves in the horizontal direction (Fig. \ref{Figure_9}(a)), leads to create only one critical area by compression, located on the right side of the last cavity
(Fig. \ref{Figure_9}(b)). An increase of the shear stresses in the vertical direction is induced (Fig. \ref{Figure_9}(c))
and two symmetric critical areas by shear stresses are generated in this direction (Fig. \ref{Figure_9}(d)). Cracks are then expected in these zones, perpendicular to the formation direction of the cavities, as experimentally observed \cite{Zimmermann2016}.

Thus, by only changing one parameter, here the dynamics of energy deposition, the interactions between the generated cavities can be controlled and tuned in order to obtain the desired designed structures.

\section{CONCLUSION}
\label{conclusion}
An application of a developed numerical tool, including a thermo-elasto-plastic model, has been carried out to study the formation of complex nano-structures, as multiple-cavity, induced by femtosecond laser tightly focused in the bulk of fused silica.
First, the case of a single deposition has been presented to explain the dynamics of a cavity formation. The evolution of the cavity radius, closely linked to the surrounded solid response, is defined by
mainly three steps. The evaluation of the principal stress in compression and the maximum shear stress, during the cavity formation,
has shown that cracks may only be induced by compression during the first step. The shear stress is not high enough in this case, due to the spherical symmetry of the system.

The case of a simultaneous double deposition
has then been investigated. It turns out that the collision of the two shock waves between the cavities leads to a daughter wave, with a larger amplitude in the perpendicular direction. The latter induces
an inhomogeneous distribution 
of compressive stresses and two elongated cavities, surrounded by an inhomogeneous overdense region. Within these conditions, due to the formation of daughter waves, the shear stress is high enough to induce critical areas where cracks may form. 
Furthermore, the 
principal stress in compression is larger than the strenght limit in compression of the material during a longer time. This behavior is observed as soon as the distance between the two energy deposition areas is not too close or too far.

The multiple-cavity case is also considered by simulating four energy depositions. First, the four energy depositions are simultaneous. That leads to the interaction
and collision of four shock waves generating daughter waves. Within these conditions, four slightly similar elongated cavities are formed, surrounded by an inhomogeneous overdense region. As in the previous case, high compressive stresses
are induced in the perpendicular direction to the energy depositions line, inducing two symmetric critical areas by compression in this direction and two symmetric critical areas by shear in the perpendicular direction. 
Then, the four energy depositions are delayed. This regime has
shown that the shock waves interact mainly in the direction of alignment of the energy depositions with an accumulation of compressive stresses in this direction. Four cavities are then generated, with slightly different sizes and with 
a different aspect ratios due to the increase
of stress (accumulation of compressive stresses) around each new cavity. The high compressive stresses accumulated induces only one critical area by compression in the front of the scanning direction.
An increase of the shear stresses in the perpendicular direction 
leads to two symmetric critical areas. Thus, depending on the energy deposition configuration, the location of critical areas changes significantly.

The results obtain in this study can be generalized to other dielectric materials. They show the importance of the choice of energy deposition setup to obtain a desired structuration of a dielectric material. Depending on the configuration, 
elongated or spherical cavities can be formed, while cracks may appear in different localized areas. They highlight also the strong potential of a numerical tool including hydrodynamic and thermo-elasto-plastic material responses, 
to predict and design laser-induced nano-structures.

\section{ACKNOWLEDGMENTS}
Vladimir Tikhonchuk and Jocelain Trela are acknowledged for fruitful discussions. The CEA and the R\'egion Aquitaine (MOTIF project) are acknowledged for supporting this work. 



\end{document}